\documentclass[amssymb,amsmath,aps,showpacs,floatfix,nofootinbib,showpacs,12pt]{revtex4}
\usepackage{epsfig}
\usepackage{color}
\usepackage{graphicx}

\def\beq{\begin{equation}}
\def\eeq{\end{equation}}

\def\be{\begin{equation}}
\def\ee{\end{equation}}
\def\bea{\begin{eqnarray}}
\def\eea{\end{eqnarray}}

\newcommand{\gsim}{\lower.7ex\hbox{$\;\stackrel{\textstyle>}{\sim}\;$}}
\newcommand{\lsim}{\lower.7ex\hbox{$\;\stackrel{\textstyle<}{\sim}\;$}}

\begin{document}

\hspace{4.5in}IPMU 10-0040

\bigskip

\title{Unitarity Boomerang}

\author{Paul H. Frampton$^{1,2}$ and Xiao-Gang He$^{1,3}$}
\affiliation{
$^1$Institute for the Physics and Mathematics of the Universe,
University of Tokyo, Kashiwa, Chiba 277-8568, JAPAN\\
$^2$Department of Physics and Astronomy,
University of North Carolina, Chapel Hill, NC 27599-3255, USA\\
$^3$Department of Physics and Center for Theoretical Sciences, National Taiwan University, Taipei, Taiwan}

\begin{abstract}
For the three family quark flavor mixing, the best parametrization is the original Kobayashi-Maskawa matrix, $V_{KM}$,
with four real parameters:
three rotation angles $\theta_{1,2,3}$ and one phase $\delta$. A popular way of presentation is by the
unitarity triangle which, however, explicitly displays only three, not four, independent parameters.
Here we propose an alternative presentation which displays
simultaneously all four parameters: the unitarity boomerang.
\end{abstract}


\maketitle

\noindent
{\bf Introduction}

\bigskip

As is well known, there are different ways of parameterizing the Kobayashi-Maskawa\cite{KM} quark mixing matrix,
$V_{KM}$. For three generations of quarks,
$V_{KM}$ is a $3\times 3$ unitary mixing matrix with three rotation angles $(\theta_1,\;\theta_2,\;\theta_3)$
and one CP violating
phase $\delta$. The magnitudes of the elements $V_{ij}$ of $V_{KM}$
are physical quantities which do not depend on parametrization. However, the value of $\delta$ does.
For example, in the Particle Data Group (PDG) parametrization \cite{PDG}, adopted from Ref.\cite{ck}, $\delta \sim 70^\circ$,
whereas the phase in the original KM parametrization
has a different value, $\delta \sim 90^\circ$. Care must be exercised in quoting a value of
$\delta$, as it depends on how the matrix is parameterized.
For example, the statement made after Eq. (11.3) in the
current edition of PDG is misleading, because it identifies,
incorrectly, the phase $\delta$ of Ref.\cite{KM}.

\bigskip

\bigskip

\noindent
It can therefore be more useful to employ only physically-measurable quantities. To this end, it has long ago
been suggested that
a unitarity triangle (UT) be used\cite{BJ} as a useful presentation for
the quark flavor mixing, especially of CP violation\cite{CCFT}.
Because of the unitary nature of the KM matrix, one has $\sum_i V_{ij}V^*_{ik} = \delta_{jk}$ and $\sum_i V_{ji}V^*_{ki} = \delta_{jk}$, where the first and second indices of $V_{ij}$
take the values $u,\;c,\;t,\;...$ and $d,\;s,\;b,\;...$, respectively.
For three generations of quarks, when $j\neq k$, these equations form closed triangles in a plane, the $UTs$.
Six $UTs$ can be formed with all of them having the same area. $A(UT)$, which is equal to half of the value of the
Jarlskog determinant \cite{jarlskog} $J$, so that $A(UT)=\frac{1}{2} J$.
The inner angles of a given $UT$ are therefore closely related
to the CP violating measure $J$. When the inner angles are measured independently,
their sum, whether it turns out to be consistent with precisely
$180^\circ$, provides a test for the
unitarity of the KM matrix. The unitarity triangle is also a popular
way, to present CP violation, with three generations of quarks.

\bigskip

\noindent
A $UT$, however, does not contain all the information encoded in the KM matrix, $V_{KM}$.
Although a $UT$ has three inner angles and three sides,
it contains only three independent parameters. The three parameters
can be chosen to be two of the three inner angles and the area, or the three sides, or some combination
thereof.
One needs an additional parameter fully to represent the physics:
this is hardly surprising, as the original
$UT$ idea of \cite{BJ} involved only two, of the three, rows
or columns of the $3 \times 3$ matrix, $V_{KM}$,

\bigskip

\bigskip

\noindent
An improved presentation is thus rendered desirable, in order better
to present the KM matrix, $V_{KM}$, diagrammatically.
In this Letter, we propose such a new diagram, the unitarity boomerang.

\bigskip

\noindent
The unitarity boomerang contains information from
a pair of UTs. The different ways of choosing the
pair contain, of course, equivalent information.
Nevertheless, the specific
choice, in the next section, was made
judiciously\cite{FH2},
such as to maximize the minimum vertex angle in
the unitarity boomerang.  This choice
is, we believe, the most convenient.

\bigskip

\noindent
{\bf Unitarity Boomerang}

\bigskip

\noindent
We indicate the KM matrix and its elements by $V_{KM} = (V_{KM})_{ij}$, with
$i = u, c, t$ and $j = b, s, d$. The unitarity of this matrix
implies $\Sigma_i V_{ij} V^*_{ik} = \delta_{jk}$ and
$\Sigma_j V_{ij} V^*_{kj} = \delta_{ik}$.
The $j \neq k$ and $i \neq k$ cases form, respectively, the
six possible different $UT$ presentations for $V_{KM}$
in a convenient two-dimensional plane.
There are, thus, a total of 18 inner angles in the six $UTs$. However, only 9
are different because, by Euclidean geometry,
each angle, in any particular $UT$, must have its equal counterpart in
another, different, $UT$.  This coincides with the fact that there are 9 different phase
expressions of the KM matrix for different parameterizations~\cite{xing2}.  To understand this simple but crucial discussion
consider the two $UTs$ defined by
\begin{eqnarray}
&UT(a)& ~~~~~ (V_{KM})_{ud}(V_{KM})^*_{ub} + (V_{KM})_{cd}(V_{KM})^*_{cb} + (V_{KM})_{td}(V_{KM})_{tb}^*  =  0
\nonumber \\
&UT(b)& ~~~~~ (V_{KM})_{ud}(V_{KM})^*_{td} + (V_{KM})_{us}(V_{KM})^*_{ts} + (V_{KM})_{ub}(V_{KM})^*_{tb}   =  0
\label{define}
\end{eqnarray}
The inner angles defined by UT (a), in Eq. (\ref{define}), are
\begin{eqnarray}
\phi_1(\beta) &=& \arg \left (-{(V_{KM})_{cd}(V_{KM})^*_{cb}
\over (V_{KM})_{td}(V_{KM})^*_{tb}}\right ) \nonumber \\
\phi_2(\alpha) &=&
\arg \left (-{(V_{KM})_{td}(V_{KM})_{tb}^* \over (V_{KM})_{ud}(V_{KM})^*_{ub}}\right ) \nonumber \\
\phi_3(\gamma) &=& \arg \left (-{(V_{KM})_{ud}(V_{KM})^*_{ub} \over (V_{KM})_{cd}(V_{KM})_{cb}^*}\right )
\label{UTa}
\end{eqnarray}
Correspondingly, the unitarity triangle, $UT (b)$ in Eq. (\ref{define}), defines another three inner angles
\begin{eqnarray}
\phi'_1(\beta')  &=& \arg \left (-{ (V_{KM})_{us}(V_{KM})^*_{ts}\over (V_{KM})_{ub}(V_{KM})^*_{tb}}\right ) \nonumber \\
\phi'_2(\alpha') &=& \arg \left ( -{(V_{KM})_{ub}(V_{KM})^*_{tb} \over (V_{KM})_{ud}(V_{KM})^*_{td}}\right ) \nonumber \\
\phi'_3(\gamma') &=& \arg \left ( -{(V_{KM})_{ud}(V_{KM})^*_{td} \over (V_{KM})_{us}(V_{KM})_{ts}^*}\right )
\label{UTb}
\end{eqnarray}
It is clear that $\phi'_2=\phi_2$.

\bigskip

\noindent
Since all the six $UTs$ have the same area $J/2$, not all the different 9 angles are independent. For example $J = |(V_{KM})_{td}(V_{KM})^*_{tb}||(V_{KM})_{ud}(V_{KM})^*_{ub}| \sin\phi_2
= |(V_{KM})_{td}(V_{KM})^*_{tb}||(V_{KM})_{cd}(V_{KM})^*_{cb}|\sin\phi_1 = |(V_{KM})_{us}(V_{KM})^*_{ts}||(V_{KM})_{ub}(V_{KM})^*_{tb}|\sin\phi'_1 = |(V_{KM})_{ud}(V_{KM})^*_{td}|| (V_{KM})_{us}(V_{KM})_{ts}^*|\sin\phi'_3$. It can be shown that only 4 independent parameters are needed to
parameterize the six $UTs$, and two different $UTs$
contain the needed 4 parameters.

\bigskip

\noindent
The values for the angles in $UT (a)$, of Eq.(\ref{define}), derived from various experiments given by PDG are\cite{PDG}:
$\phi_1 = (21.46\pm 0.98)^\circ$
(derived from data on $\sin(2\phi_1) = 0.681\pm 0.025$), and
the values for $\phi_2$ and $\phi_3$ are  $(88^{+6}_{-5})^\circ$ and $(77^{+30}_{-32})^\circ$, respectively.
These values are consistent with the unitarity of the KM matrix within error bars,
and therefore also with a choice of presentation
which we now formulate in terms of a novel combination of
two different unitarity triangles (a) and (b).
$UT(a)$, defined by
Eq. (\ref{define}), is almost a right triangle, by virtue of $\phi_2$.
Numerically, the angles $\phi'_1$ and $\phi'_3$ are close to $\phi_1$ and $\phi_2$, respectively.
All the angles in the two $UTs$ are sizable, making experimental determination of them merely challenging, while for the other
four choices of $UT$ there is always, at least, one small angle where measurement may be exceptionally difficult.
It is therefore easiest to work with the two $UTs$, $UT(a)$ and $UT(b)$,
for practical purposes. We now show that, by combining information from these two $UTs$, into the boomerang diagram
\footnote{The name arises from resemblance to the hunting instrument.}
 displayed in Fig. 1,
all information needed to specify the KM matrix, $V_{KM}$, can be extracted.

\bigskip

\begin{figure}[hb]
\includegraphics[width=3in]{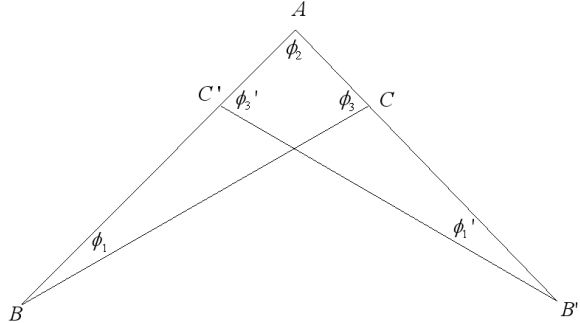}
\caption{The unitarity boomerang. The sides are: $AC = |(V_{KM})_{ud}(V_{KM})^*_{ub}|$, $AC' = |(V_{KM})_{ub}(V_{KM})^*_{tb}|$,
$AB = |(V_{KM})_{td}(V_{KM})^*_{tb}|$,
$AB' = |(V_{KM})_{ud}(V_{KM})^*_{td}|$, $BC = |(V_{KM})_{cd}(V_{KM})^*_{cb}|$ and $B'C' = |(V_{KM})_{us}(V_{KM})^*_{ts}|$.
\label{boomerang}}
\end{figure}

\bigskip

\noindent
The unitarity boomerang is formed by locating the common angle $\phi'_2 = \phi_2$ from the two $UTs$ of $UT (a)$ and $UT (b)$ at the top point $A$ and the shortest
sides, $AC = |(V_{KM})_{ud}(V_{KM})^*_{ub}|$ and $AC' = |(V_{KM})_{ub}(V_{KM})^*_{tb}|$, on the opposite sides.
The other sides are: $AB = |(V_{KM})_{td}(V_{KM})^*_{tb}|$, $AB' = |(V_{KM})_{ud}(V_{KM})^*_{td}|$, $BC = |(V_{KM})_{cd}(V_{KM})^*_{cb}|$ and $B'C' = |(V_{KM})_{us}(V_{KM})^*_{ts}|$.
We emphasize that Fig. (\ref{boomerang}) is drawn with the central experimental values of

\noindent
$AC =  3.50\times 10^{-3}$ , $AC' =  3.59\times 10^{-3}$,  $AB =  8.73\times 10^{-3}$,
$AB' =  8.51\times 10^{-3}$, $BC =  9.36\times 10^{-3}$ and $B'C' =  9.19\times 10^{-3}$.

\bigskip

\noindent One can choose the area ($J/2$) of the triangles, two inner angles from one of the $UTs$ (for example $\phi_1$ and $\phi_2$),
and a third angle from the other $UT$ (for example
$\phi'_3$) as the four independent parameters.

\bigskip

\noindent
{\bf Original KM parametrization and Unitarity Boomerang}

\bigskip

To show explicitly how the unitarity boomerang can provide all information needed to specify the
quark flavor mixing, we work with
a specific parametrization, $V_{KM}$, originally given by Kobayashi and Maskawa\cite{KM}

\begin{eqnarray}
V_{KM} = \left ( \begin{array}{ccc} c_1& - s_1 c_3& - s_1 s_3\\s_1c_2&c_1c_2c_3 - s_2s_3 e^{i\delta}&c_1c_2s_3 + s_2c_3 e^{i\delta}\\
s_1s_2&c_1s_2c_3 + c_2 s_3 e^{i\delta}& c_1s_2 s_3 - c_2c_3 e^{i\delta}\end{array}
\right )\;.
\end{eqnarray}
One can also work with other parameterizations, such as that adopted by the PDG.
But we find an interesting feature of the original KM parametrization which turns out to be very convenient for the discussions of
the unitarity boomerang.

\bigskip

\noindent
Using experimental values\cite{PDG} for for $(V_{KM})_{us} = 0.2257\pm 0.0010$, $(V_{KM})_{ub} = 0.00359\pm 0.00016$ and
$(V_{KM})_{td} =0.00874^{+0.00026}_{-0.00037}$, one finds that $s_2 s_3 <<1$. At a
few percent level, one has $(V_{KM})_{tb} = (c_1s_2s_3 - c_2c_3 e^{-i\delta}) \approx - c_2c_3 e^{-i\delta}$.

\bigskip

\noindent
Then

\begin{eqnarray}
\phi_2 & = & \arg(-{ s_1 s_2*(c_1s_2s_3 - c_2c_3 e^{-i\delta})\over c_1*(- s_1s_3)}) \nonumber \\
& \approx &  \arg({s_1 s_2*(- c_2c_3 e^{-i\delta})\over c_1* s_1s_3 }) = \pi - \delta.
\label{re-delta}
\end{eqnarray}
The CP violating phase $\delta$, in this parametrization, is equal to $\pi - \phi_2$, to a good approximation~\cite{koide2}.

\bigskip

\noindent
The fact that $\phi_2 = (88^{+6}_{-5})^\circ$ implies $\delta \approx 90^\circ$.
The approximate right angle at the top
of the boomerang diagram may indicate that CP, from a deeper perspective, is maximally
violated\cite{fx,koide}.
Kobayashi and Maskawa, with remarkable prescience, made an excellent choice of parametrization.
We suggest that the original parametrization of Kobayashi-Maskawa matrix be used as the standard parametrization.
A parametrization suggested by Fritzsch and Xing\cite{fx}, which also has its phase close to $\phi_2$,
is another alternative interesting parametrization.
From the unitarity boomerang, one can easily obtain approximation solutions for the
four physical parameters. One first notices that the relation in Eq.(\ref{re-delta})
allows one to read off the $\delta$ from the top angle in the diagram.
Taking the ratio, of the two sides $AC/AC'$ or $AB/AB'$, one obtains $|(V_{KM})_{ud}/(V_{KM})_{tb}^*| \approx c_1$
since $|(V_{KM})_{tb}|$ is very close to 1.
With $c_1$ and therefore $s_1$ known, the length of the sides AB and AC' then provide the values for $s_2$ and $s_3$.

\bigskip

\noindent
One can obtain more precise solutions by using the following information from four sides,
$AC = a$, $BC = b$, $AB = c$ and $AB' = d$ of the unitarity boomerang:
\begin{eqnarray}
&&a = |(V_{KM})_{ud}(V_{KM})^*_{ub}| = c_1s_1s_3\;,\;\;b = |(V_{KM})_{cd}(V_{KM})^*_{cb}|
= s_1c_2|c_1c_2s_3+s_2c_3e^{-i\delta}|\;,\nonumber\\
&&c = |(V_{KM})_{td}V_{KM})^*_{tb}| = s_1s_2|c_1s_2s_3-c_2c_3 e^{-i\delta}|\;,\;\;d =
|(V_{KM})_{ud}(V_{KM})^*_{td}|= c_1 s_1 s_2\;.
\end{eqnarray}

\noindent
Using the above, one can express $s_{1,2,3}$ and $\delta$ as functions of $a$, $b$,
$c$ and $d$. The KM parameters can be determined. For example
\begin{eqnarray}
a^2- c_1^2 + c^4_1 \left ( {c^2\over d^2} - {b^2\over c^4_1 -c_1^2 +d^2}\right ) = 0\;.
\label{solution}
\end{eqnarray}
Solving for the roots of the above equations, the $c_1^2$ is determined up to four
possible discrete solutions. Restricting to real positive solutions with magnitude less than 1, one can further limit the choices.

\bigskip

\noindent
The other angles, and the phase, can be determined from the following relations
\begin{eqnarray}
&&s_2 = {d\over c_1 s_1}\;,\;\;s_3 = {a\over c_1 s_2}\;,\nonumber\\
&&\cos\delta = {b^2/s_1^2 c_2^2-(c_1^2c_2^2s_3^2 + s^2_2 c_3^2)\over 2  c_1 c_2 s_2 c_3 s_3}
= {c^2_1 s^2_2 s^2_3 + c_2^2 c_3^2 - c^2/s_1^2s_2^2\over 2 c_1 c_2 s_2 c_3 s_3}\;.
\end{eqnarray}
After applying the constraint on $c^2_{2,3}$, that they satisfy $0 \leq c^2_{2,3} \leq 1$,
the solution is even more restricted. Putting in numerical values, for the sides, and
comparing with the approximate solution above, we find that a unique solution survives.

\bigskip

\noindent
Numerically, with the current central values for $a$, $b$ $c$ and $d$, we obtain
\begin{eqnarray}
c_1 = 0.97419\;,\;\;s_2 = 0.0387\;,\;\;s_3 = 0.0162\;,\;\;\delta = 88.83^\circ\;.
\end{eqnarray}
and these numbers are self consistent.

\bigskip

\noindent
One should be aware, that there remain errors, on the sides and angles of the boomerang.
This leads to distortion of the $UB$ away from the true one. When constructing the $UB$, one can first use measurable quantities
without assuming unitarity to form one of the $UT$, say, the $UT$ defined by triangle $ABC$ in Fig. 1. This can be achieved by using the
measured $\alpha$ and $\beta$ and also the length of side $AB$, $c = |(V_{KM})_{td}(V_{KM})_{tb}^*|$.
The major error comes from the
uncertainty in $|(V_{KM})_{td}(V_{KM})_{tb}^*|$ measured from $B_b - \bar B_d$ mixing. Assuming $|(V_{KM})_{tb}|$ is almost one, then~\cite{PDG},
$|(V_{KM})_{td}| = (8.09\pm 0.6)\times 10^{-3}$.
One then uses information on the values of
$|(V_{KM})_{ud}|$ and $|(V_{KM})_{ub}|$ to construct the sides $AB'$ and $AC'$ to complete the boomerang.
The error in $|(V_{KM})_{td}|$ will cause uncertainty in the side $AB'$ of the $UB$ with $d = (7.88\pm0.58)\times 10^{-3}$. At present
within error bars, one cannot be sure which side, $AB$ or $AB'$, is longer.
Further reduce the errors in $|(V_{KM})_{td}(V_{KM})^*_{tb}|$ can be achieved by better understanding of the bag factor in $B_d - \bar B_d$ mixing~\cite{PDG}.
Another way to improve the situation is to note that the value
$|(V_{KM})_{tb}|/|(V_{KM})_{ud}|$ plays an important role which also determine the ratio of $AC$ and $AC'$.
Therefore precise measurement of $|(V_{KM})_{tb}|$ is crucial in constructing an accurate $UB$.
Future studies of top quark decay and single top quark production at colliders, such as the LHC, will provide useful information.

\bigskip

\noindent
To give a quantitative feeling, we have carried out an estimate assuming that
the errors in $a$, $b$, $c$ and $d$ are given by the current PDG data with Gaussian errors to obtain the resultant errors in the KM angles.
We obtain
 $\Delta c_1 = 0.046$, an error which is reasonably small.
But errors on $s_{2,3}$ are large with
 $\Delta s_2 = 0.032$ and $\Delta s_3 = 0.077$. Such a larger error
 bolsters preference for the boomerang, to disentangle,
most perspicuously, the quark flavor mixing. Note that
errors, on $s_{2,3}$, are due to empirically-generated uncertainties on
$(V_{KM})_{td}$, $(V_{KM})_{cb}$ and $(V_{KM})_{ub}$.

\bigskip

\noindent
Indeed, when we look more closely at Eq. (\ref{solution}),
it does turn out that the quantity $c$ enters that equation, only in a combination
($c^2/d^2$), just so that $(V_{KM})_{td}$ cancels out. If one takes into account,
the errors are
reduced to $\Delta c_1 = 0.032$, $\Delta s_2 = 0.023$ and $\Delta s_3 = 0.055$.

\bigskip

\noindent
If uncertainties on all four sides can be
reduced, say by another factor of three, we project that errors can be
reduced to $\Delta c_1 = 0.011$, $\Delta s_2 = 0.076$ and $\Delta s_3 = 0.018$,
thus illustrating how the chosen
boomerang may, in the foreseeable future,
return to increase human knowledge. Our proposal, to move from a single triangle to a boomerang combination,
therefore reflects, more than anything else,
the increase in precision which is justifiably anticipated from the high-energy experiments.

\bigskip

\noindent
{\bf Discussion}

\bigskip

\noindent
The most popular way to present the flavor mixing for three generations of quarks is by a
unitarity triangle which, however, explicitly displays only three of four
independent parameters. To have a diagrammatical representation for the full four independent
parameters, we have proposed improvement to the unitarity boomerang.

\bigskip

\noindent
By studying the unitarity boomerang, one can obtain all the information enshrined in KM matrix. 
We find that the original parametrization by Kobayashi and Maskawa is particularly convenient 
for this purpose. The angle $\phi_2$ in the boomerang diagram, to a good approximation, 
can be identified with the phase $\delta$ in the original KM parametrization \cite{KM}.
The fact that $\phi_2 = (88^{+6}_{-5})^\circ$ implies $\delta \approx 90^\circ$, so that this parametrization
may be the right one to study assiduously, in order to probe further the connection to the
origin of, possibly maximal, CP violation.
We, therefore, humbly submit that the original parametrization of KM matrix be kept
as the standard, and that the unitarity boomerang
shown in FIG.1 be used unambiguously to present
the experimental information.

\bigskip

\noindent
{\bf Acknowledgements}

\bigskip

\noindent
We thank Lu-Hsing Tsai for helps in preparation of the figure. This work was supported by the World Premier
International Research Initiative
(WPI initiative) MEXT, Japan. The work of P.H.F. was also supported by U.S. Department
of Energy Grant No. DE-FG02-05ER41418. The work of X.G.H was supported by the NSC and NCTS
of ROC.

\end{document}